\documentclass[twocolumn,prl,showpacs,amsmath,amssymb,floatfix,superscriptaddress]{revtex4-2}

\usepackage[latin9]{inputenc}
\usepackage{graphicx}
\usepackage[normalem]{ulem}
\usepackage{esint}
\usepackage{chemformula}
\usepackage{multirow}
\usepackage{makecell}
\usepackage{booktabs}
\usepackage[normalem]{ulem}
\usepackage{siunitx}
\usepackage{nameref}
\newcommand{\AVQO}{AV$_{2}$Q$_{2}$O }

\usepackage{ragged2e}
\usepackage{caption}
\usepackage[linktocpage=true,
  colorlinks=true, 
  pdfborder={0 0 0}, 
  linkcolor=blue,
  citecolor=red,
  filecolor=yellow,
  urlcolor=blue,
  bookmarks,
  pdfauthor={},
]{hyperref}

\makeatletter

\usepackage{dcolumn}%Align table columns on decimal point
\usepackage{bm}% bold math
\ifdefined\showcaptionsetup
 \PassOptionsToPackage{caption=false}{subfig}
\fi
\usepackage{subfig}
\makeatother

\begin{document}
\title{Is altermagnetism in vanadium oxychalcogenides a lost cause?}

% Authors
\author{Bishal Thapa}
\email{bthapa3@gmu.edu}
\affiliation{George Mason University, Department of Physics and Astronomy} \affiliation{Quantum Science and Engineering Center,  Fairfax, USA}

\author{Po-Hao Chang}
\email{pchang8@gmu.edu}
\affiliation{George Mason University, Department of Physics and Astronomy} \affiliation{Quantum Science and Engineering Center,  Fairfax, USA}

\author{Kirill Belashchenko}
\email{belashchenko@unl.edu}
\affiliation{Department of Physics and Astronomy and Nebraska Center for Materials and Nanoscience,
University of Nebraska-Lincoln, Lincoln, Nebraska 68588, USA}

\author{Igor I. Mazin}
\email{imazin2@gmu.edu}
\affiliation{George Mason University, Department of Physics and Astronomy} \affiliation{Quantum Science and Engineering Center,  Fairfax, USA}

\begin{abstract}
Vanadium-based oxychalcogenide compounds with the inverse Lieb-lattice (ILL) structural pattern have been recently proposed as candidate altermagnets (AM).
However, early studies postulated ferromagnetic interlayer coupling, a critical requirement for preserving the bulk AM state.  
Here we present a systematic survey of the complete AV$_2$Q$_2$O family ($\text{A} = \text{K, Rb, Cs}$; $\text{Q} = \text{S, Se, Te}$) in terms of their magnetic ordering and interlayer coupling. 
While \emph{intralayer} exchange interaction favors AM ordering in a single ILL layer across the entire family, the relatively weak \emph{interlayer} coupling in most cases favors Kramers-degenerate antiferromagnetic order with a doubled magnetic unit cell. This means that most stoichiometric bulk materials, including the previously proposed candidate \ch{KV_2Se_2O}, are not altermagnetic, with \ch{CsV_2Te_2O} being the only exception.  
Using hole doping to simulate alkali vacancies, we show that realistic deviations from stoichiometry to not change the magnetic ground state in these compounds.
\end{abstract}

\maketitle

\paragraph{Introduction ---}
Altermagnetism (AM) has emerged as a  frontier in contemporary
condensed matter physics \cite{mazin_editorial_2022,smejkal_beyond_2022,mazin_prediction_2021,smejkal_anomalous_2022,smejkal_giant_2022,smejkal_emerging_2022,bai2024altermagnetism} and spintronics, offering the prospect of
generating large spin splittings of nonrelativistic origin on the
order of electronvolts {in the absence of net magnetization} \cite{zhang2025giant,yang2025large}. This remarkable property positions altermagnetic
materials as promising candidates for next-generation spintronic applications \cite{gonzalez2021efficient,bose2022tilted,bai2022observation,shao2021spin,karube2022observation}.

Inverse Lieb lattice (ILL) compounds have attracted considerable attention
as a versatile platform for hosting altermagnetic states \cite{kaushal_altermagnetism_2024,durrnagel_altermagnetic_2024, wei_2_2025,jiang_discovery_2025,zhang_crystal-symmetry-paired_2025,chang_inverse_2025}. 
Their appeal stems from several key features: natural abundance of the constituent elements, $d$-wave altermagnetic spin-splitting symmetry, availability of metallic compounds \cite{jiang_discovery_2025,zhang_crystal-symmetry-paired_2025}, and directionally distinct second-nearest-neighbor exchange paths potentially leading to large chiral magnon splitting \cite{chang_inverse_2025}. %much closer inequivalent exchange coupling parameters $J_{2}$.
The characteristic architecture of ILL compounds --- comprising
transition-metal (TM) layers decorated by ligand atoms and separated by various
filler layers --- provides highly tunable electronic and magnetic
properties, including metallicity, Fermi surface, and N\'{e}el temperature $T_{N}$.
However, altermagnetism is not a universal feature of ILL-based materials, necessitating systematic investigation of the magnetic order in this family \cite{PhysRevB.110.054406,litvin1974spin,brinkman1966theory,chang_inverse_2025}.

We have previously studied \emph{intralayer} magnetic ordering in
a broad range of ILL systems, revealing intriguing patterns in AM
stabilization \cite{chang_inverse_2025}. The analysis of these compounds uncovered a potentially
significant correlation: transition metals {with less than half-filled $3d$ shells} exhibit stronger propensity for {  in-plane} altermagnetic
order. While a unified microscopic explanation for this trend remains
elusive (only a special case of the $3d^{5}$ \ch{La2O3Mn2Se2} compound has been discussed
in detail \cite{garcia-gassullMicroscopicOriginMagnetic2025}), this observation could provide valuable guidance for identifying
promising AM candidates.

Among the experimentally characterized AM candidates,
there are two metallic systems belonging to the ILL family, which are both
vanadium-based materials with $T_{N}$ around room temperature \cite{jiang_discovery_2025,zhang_crystal-symmetry-paired_2025}.
In addition, insulating V$_{2}$Se$_{2}$O, which consists of similar ILL layers without any intercalating filler layers, has also been synthesized \cite{lin_structure_2018}.
Its magnetic ordering has not yet been determined experimentally, but
theoretical calculations suggest an AM ground
state \cite{yu_high_2021,ma_multifunctional_2021}. These observations collectively suggest that vanadium-based ILL compounds are prime candidates for the technologically desirable metallic $d$-wave AM \cite{chang_inverse_2025}.

However, AM of bulk ILL compounds hinges on a critical issue that has not been systematically addressed:
the ILL layers must be stacked ferromagnetically along the $c$ axis; otherwise, the magnetic unit cell is doubled, and the compound becomes a conventional collinear antiferromagnet with Kramers-degenerate band structure.
Many earlier theoretical analyses have either focused on hypothetical monolayer systems or assumed ferromagnetic interlayer coupling
\cite{ablimit_weak_2018,jiang_discovery_2025,zhu_multipiezo_2024,zou_monolayer_2024}, thus leaving the AM character unresolved for the existing material candidates.

In this paper, guided by the available experimental data \cite{kelly2023structure,valldor2016bad}, we present a systematic survey
of the magnetic properties across the complete family of metallic vanadium-based
ILL compounds with the general formula AV$_{2}$Q$_{2}$O, where A
= K, Rb, Cs and Q = S, Se, Te, focusing on interlayer exchange coupling. 
This family represents an ideal testbed for several reasons. First,
according to our earlier analysis \cite{chang_inverse_2025}, vanadium with its less than half-filled $3d$ shell should form ILL layers with \emph{intralayer} AM, and we
can systematically explore the effect of the alkali metal
filler (A) and chalcogen ligand (Q) on the \emph{interlayer} magnetic ordering. %metallicity, and N\'{e}el temperature.
Second, they are all metallic. Third, several of these compounds have already been synthesized and structurally characterized, making the whole family experimentally accessible.

In agreement with our earlier results \cite{chang_inverse_2025}, the intralayer 
exchange coupling strongly favors monolayer AM in all nine compounds. However, we find that much weaker interlayer exchange coupling favors antiferromagnetic (AF) stacking in all cases except \ch{CsV2Te2O}, excluding AM order in all but this one (hypothetical) compound. This conclusion is in contradiction with the original claim that KV$_{2}$Se$_{2}$O is AM~\cite{jiang_discovery_2025} but in agreement with the recent neutron diffraction experiment \cite{KVSeOneutron}.
We have also checked that this conclusion is robust with respect to a reasonable charge doping, which represents deviations from stoichiometry and incomplete population of the alkali-atom sited in the intercalating layers, such as Rb layers in \ch{Rb_{$1-\delta$}V2Te2O} \cite{zhang_crystal-symmetry-paired_2025} .

The rest of this paper is organized as follows. Section II briefly describes the crystal and magnetic structure of the AV\textsubscript{2}Q\textsubscript{2}O
compounds. Section III details the computational methodology. Sections IV presents the main results for the exchange coupling in both stoichiometric and hole-doped compounds. 
Section V concludes the paper.

\paragraph{Crystal and magnetic structure ---\label{sec:structure}}

\begin{figure}[h]
\centerline{\includegraphics[width=\columnwidth,clip]{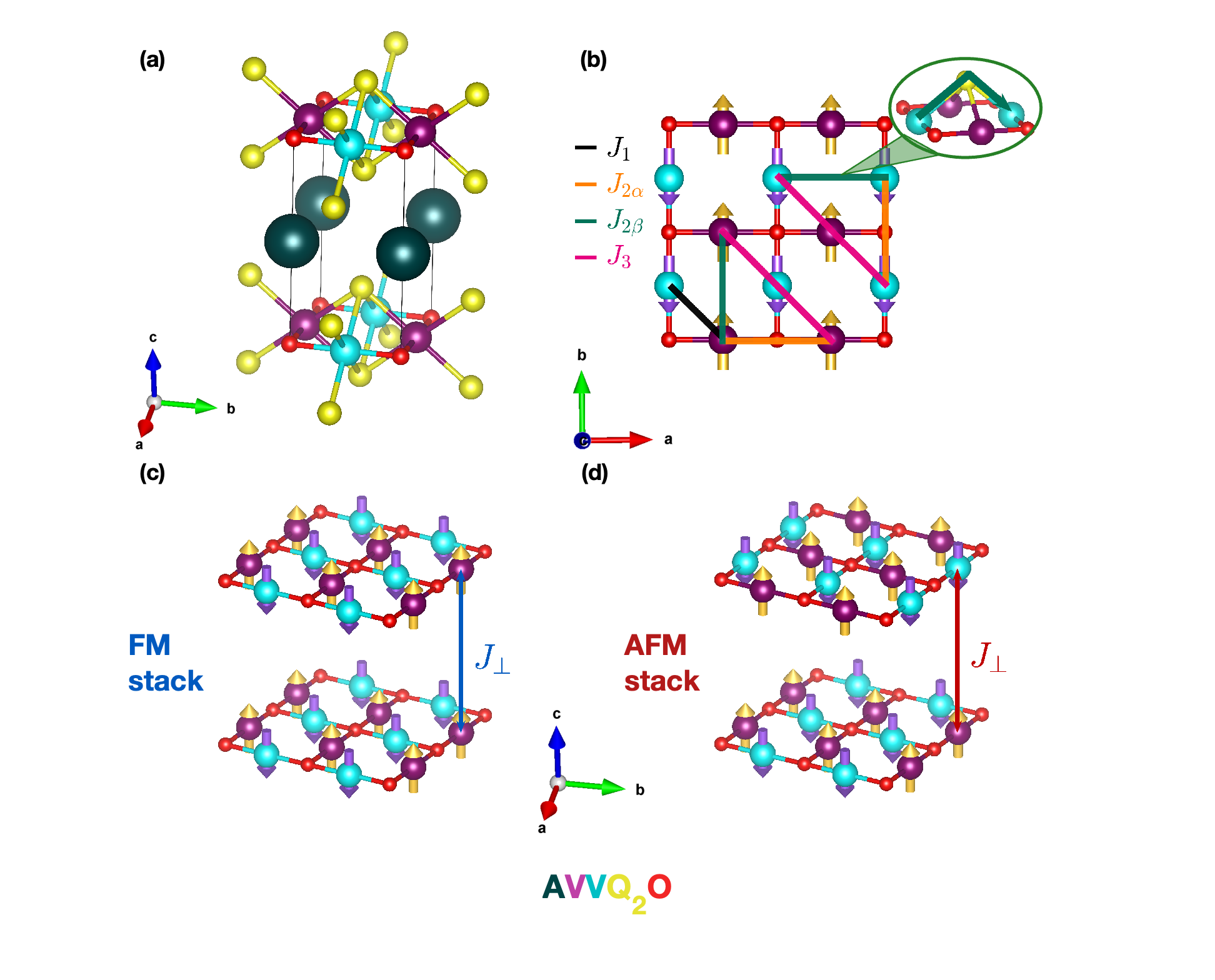}}
\caption{\justifying (a) Crystal structure of \AVQO showing octahedra forming prismatic slabs separated by the intercalating A layers. (b) Magnetic lattice viewed along the $c$ axis. Cyan and magenta colors denote V atoms on two magnetic sublattices. $J_1$, $J_{2\alpha}$, $J_{2\beta}$, $J_3$: exchange parameters.
(c) AM bulk ordering: all ILL layers are stacked ferromagnetically.
(d) AF bulk ordering: ILL layers are stacked antiferromagnetically. A and Q atoms are omitted for clarify in panels (c) and (d).
$J_{\perp}$: interlayer magnetic coupling.
}
\label{fig:Structure}
\end{figure}

The compounds in the isoelectronic $\mathrm{AV_2Q_2O}$ family (A = K, Rb, Cs; Q = S, Se, Te) crystallize in the primitive tetragonal space group $P4/mmm$ (No.~123), with the experimental lattice constants are reported in Ref.~\cite{kelly2023structure}. The A, V, Q, and O atoms atoms occupy the $1b$, $2f$, $2h$, and $1a$ Wyckoff sites, respectively.
This structure consists of well-separated $\mathrm{V_2Q_2O}$ layers stacked without a lateral shift along the crystallographic $c$ axis and interleaved by layers of alkali-metal ions. The V atoms within each layer form an ILL, as shown in Fig.~\ref{fig:Structure}.

The exchange coupling and magnetic ordering within the ILL layer were studied in detail in Ref. \onlinecite{chang_inverse_2025}. The relevant intralayer exchange parameters $J_1$, $J_{2\alpha}$, $J_{2\beta}$, and $J_3$ are indicated in Fig. \ref{fig:Structure}(b).
As explained below, in all studied compounds, these parameters reliably stabilize AM order within an isolated ILL monolayer, in agreement with the conclusions of Ref. \onlinecite{chang_inverse_2025}.
Our central focus will, therefore, be on the \emph{interlayer} interaction, denoted $J_{\perp}$ in Fig. \ref{fig:Structure}(c-d), which couples V spins in the adjacent $\mathrm{V_2Q_2O}$ layers. Instead of calculating individual pair exchange parameters connecting V atoms in such adjacent layers, we will study the total interlayer coupling using total energy calculations for FM and AF layer stackings, as shown in Fig. \ref{fig:Structure}(c) and \ref{fig:Structure}(d), respectively. A compound with ferromagnetic stacking of Fig. \ref{fig:Structure}(c) is AM, while AF stacking of Fig. \ref{fig:Structure}(d) is a conventional antiferromagnet with a doubled magnetic unit cell.

\paragraph{Computational details ---\label{sec:method}}

Most first-principles calculations were performed using the numerical-orbital-based
density functional theory (DFT) code OpenMX~\cite{2003variationallyozaki,OpenMX}.
Core electrons were treated using norm-conserving pseudopotentials
\cite{morrisonNonlocalHermitianNormconserving1993,
vanderbiltSoftSelfconsistentPseudopotentials1990}.
The Perdew--Burke--Ernzerhof (PBE) generalized gradient approximation
was employed to describe exchange--correlation effects~\cite{perdewGeneralizedGradientApproximation1996}.
Pseudoatomic basis functions and pseudopotentials for the
$\mathrm{AV_2Q_2O}$ compounds were taken from the OpenMX distribution.
For the self-consistent calculations, the energy cutoff for numerical grid integration was set to 300 Ry, and a $15\times15\times5$ Monkhorst-Pack k-point mesh was used for Brillouin zone sampling.

The in-plane exchange constants were obtained \cite{chang_inverse_2025} using the Green's function-based linear-response technique  \cite{antropovExchangeInteractionsMagnets1997} implemented in OpenMX\cite{OpenMX}, using the altermagnetic state as a reference configuration. 
On the other hand, the interlayer coupling $J_{\perp}$ was found using the total energy difference, $J_{\perp} =  \left(E_{\mathrm{FM}} - E_{\mathrm{AFM}}\right)/8$, between the AM  and AF orderings within the doubled unit cell, shown in %To this end, we employed a supercell doubled along the $c$, as shown in
Figs. ~\ref{fig:Structure}(c) and ~\ref{fig:Structure}(d). The total energies were calculated using the Vienna Ab initio Simulation Package (VASP) ~\cite{VASP-1,VASP-2,VASP-3}, within the PBE generalized gradient approximation, 
using a $24\times24\times7$ $k$-point mesh to achieve sufficient convergence. 

The results reported in the following were obtained without spin-orbit coupling; we verified for selected cases that its inclusion had a negligible effect on the relevant energy differences.
For selected materials we have also evaluated $J_{\perp}$ using the all-electron package Wien2k \cite{Wien2k}; the results were consistent with those obtained in VASP.

\paragraph{Intralayer Exchange Coupling ---}
\phantomsection
\label{sec:intralayer}

\begin{table}[htbp]
\caption{Calculated exchange parameters (meV).}
\setlength{\tabcolsep}{1pt} 
\begin{tabular}{l S S S S S S}
\toprule
Compound
& {$J_{1}$}
& {$J_{2\alpha}$}
& {$J_{2\beta}$}
& {$J_{3}$}
& {$J_{4}$}
& {$J_{\perp}$} \\
\midrule
KV$_2$S$_2$O    & 39.10 & 35.94 &  2.78 & 6.78 & -0.16 &  0.530 \\
KV$_2$Se$_2$O   & 41.56 & 34.06 &  0.84 & 7.00 & -0.16 &  0.554 \\
KV$_2$Te$_2$O   & 40.28 & 31.36 & -2.88 & 6.82 & -1.24 &  0.383 \\

RbV$_2$S$_2$O   & 39.62 & 33.20 &  4.48 & 7.00 &  0.14 &  0.350 \\
RbV$_2$Se$_2$O  & 44.80 & 26.22 & -1.42 & 7.54 & -1.10 &  0.339 \\
RbV$_2$Te$_2$O  & 40.36 & 29.94 & -2.58 & 6.92 & -1.08 &  0.197 \\

CsV$_2$S$_2$O   & 41.22 & 32.06 &  2.46 & 7.20 &  0.10 &  0.246 \\
CsV$_2$Se$_2$O  & 42.64 & 30.80 &  0.88 & 7.34 & -0.14 &  0.217 \\
CsV$_2$Te$_2$O  & 42.48 & 27.92 & -2.04 & 6.94 &  1.62 & -0.080 \\
\bottomrule
\end{tabular}
\label{tab:jex}
\end{table}

The calculated intralayer exchange coupling parameters are listed in Table \ref{tab:jex}.

\begin{figure}[h]
\centerline{\includegraphics[scale=0.5]{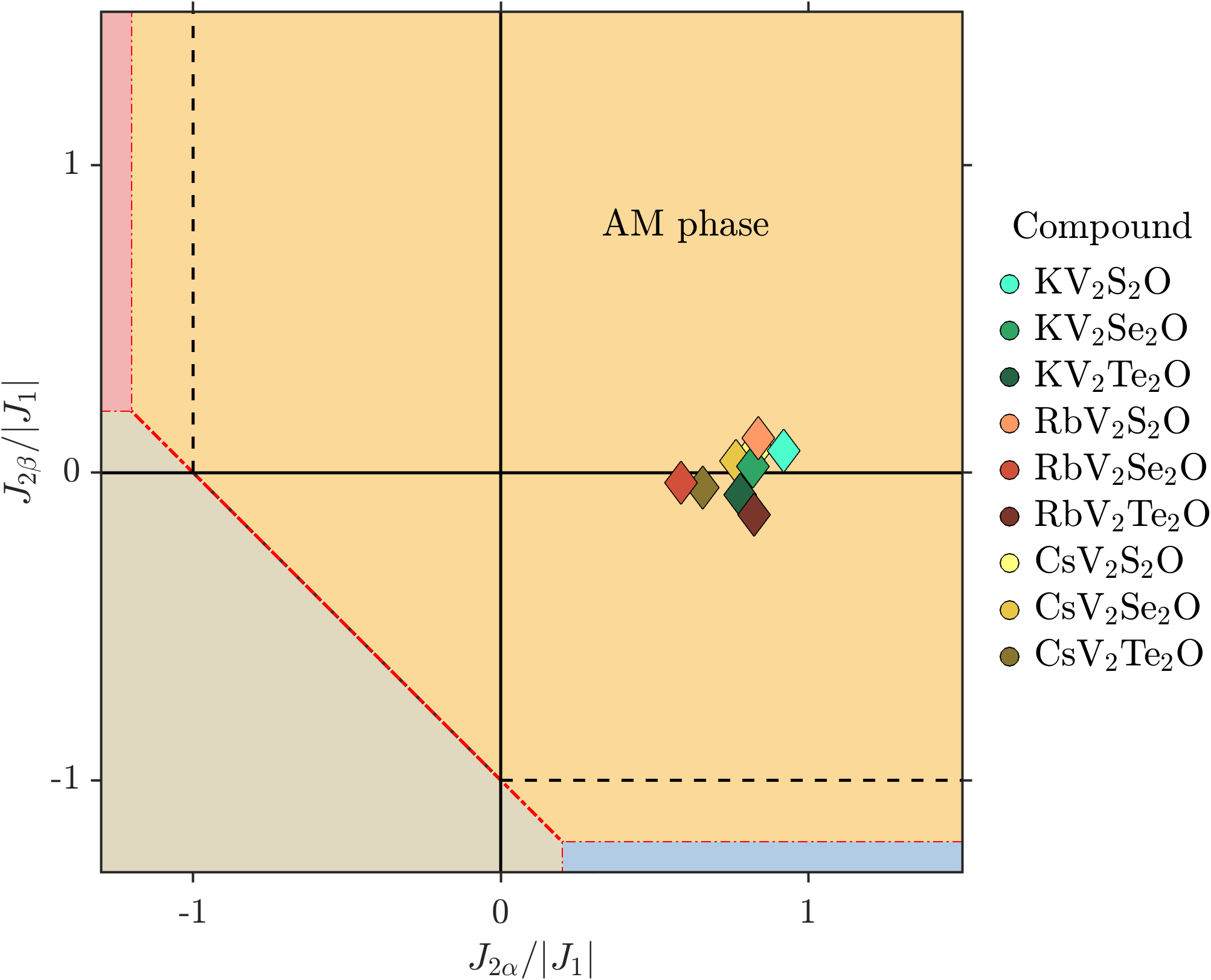}}
\caption{\justifying Magnetic phase diagram in the $(J_{2\alpha}/|J_{1}|,\,J_{2\beta}/|J_{1}|)$ parameter space. The red lines denote the phase boundaries for $J_{3}=0.1|J_1|$. For comparison, black dashed lines show the phase boundaries at $J_3=0$ (note that  $J_{3}>0$ expands the AM region). The unlabeled regions are various non-AM phases, as discussed in Ref. \onlinecite{chang_inverse_2025}. }
\label{fig-phase}
\end{figure}

As explained in Ref. \onlinecite{chang_inverse_2025}, the magnetic phase stability in an ILL layer may be inferred from the magnetic phase diagram in the $(J_{2\alpha}/|J_{1}|,J_{2\beta}/|J_{1}|)$ parameter space, which is shown in Fig. \ref{fig-phase}.

As seen from Table \ref{tab:jex}, all compounds exhibit large antiferromagnetic
$J_{1}$ and ferromagnetic $J_{2\alpha}$ of a comparable magnitude; both these strong couplings favor the AM state. 
An interesting trend observed across a wider range of compounds %in the earlier study
\cite{chang_inverse_2025} is that $J_{2\beta}$ rarely acts as the deciding factor for the in-plane ordering; it is typically either FM or weakly AF. Here, in AV$_2$Q$_2$O compounds, it is quite small, and its effect on the magnetic stability is negligible. In contrast, it is more common to find $J_{2\alpha}$ to be strongly AF, leading to $90^{\circ}$ noncollinear states as found in \ch{La2O3Co2Se2} \cite{fuwa_orthogonal_2010} and \ch{La2O3Fe2Se2} \cite{gunther_magnetic_2014,mccabe_weak_2014}. In these compounds, the continuous frustration at the Heisenberg level results in ground states that are determined by higher-order interactions.

The effect of longer-range interactions (beyond $J_{2}$) may be included as phase boundary shifts in Fig. \ref{fig-phase}. 
Table \ref{tab:jex} shows the $J_{3}$ coupling is ferromagnetic in all compounds, providing additional stabilization for the AM state. Surprisingly,
$J_{3}$ is remarkably similar across distinct compounds, suggesting that this coupling is insensitive to local chemical variations and may be
governed by more universal electronic or structural features common to the series. 
The $J_{4}$ parameter is negligibly small, indicating that more distant magnetic couplings have no effect on the magnetic ground state.
This validates the truncation of the Heisenberg model at third-nearest neighbors.
Overall, the analysis of the exchange coupling confirms that the \emph{intralayer} magnetic order in all studied AV$_2$Q$_2$O compounds is reliably AM.

\paragraph{Interlayer Coupling and the Bulk Ground State ---}While the preceding discussion of intralayer exchange coupling shows
that AM ordering is strongly favored within each layer, the three-dimensional magnetic ground state depends critically 
on the interlayer coupling $J_\perp$ between the adjacent ILL layers, which are also listed in Table \ref{tab:jex}. 
We find that nearly all compounds in this family, with the single exception of \ch{CsV2Te2O}, 
favor AF interlayer stacking.

The preference for AF interlayer stacking has profound consequences for the magnetic symmetry. Such stacking doubles the magnetic unit cell
along the $c$-axis, turning the compound into a conventional, Kramers-deneregate antiferromagnet despite satisfying the necessary intralayer conditions for altermagnetism.
In this configuration, the two spin sublattices between different layers can be mapped onto each another through a lattice translation, violating the symmetry requirements
for altermagnetism \cite{smejkal_beyond_2022}. 

Thus, while vanadium-based ILL compounds remain promising for hosting altermagnetism within single layers, bulk altermagnetism requires careful attention to interlayer ordering. Among the nine compounds considered here, only \ch{CsV2Te2O} appears to be suitable, based on our DFT calculations.

In this connection, it is interesting to revisit a related \ch{Sr2CrO2Cr2OAs2} ILL compound, which was shown in Ref. \onlinecite{chang_inverse_2025} to host robust AM order within its Cr-based ILL layers. That compound also has a high $T_N$ and is metallic with a highly anisotropic $d$-wave Fermi surface. In contrast to other known ILL-based oxychalcogenides, the ILL layers in \ch{Sr2CrO2Cr2OAs2} alternate with perovskite-like CrO$_2$ layers, which, taken individually, are not AM. Nevertheless, the bulk compound belongs to an AM spin group. We emphasize that this AM character does not depend on whether the interlayer ordering is FM or AF, because in either case the magnetic unit cell contains only one ILL layer.

\paragraph{Influence of Charge Doping on Exchange Coupling ---}

Given that most stoichiometric AV$_2$Q$_2$O compounds exhibit weak AF interlayer coupling, it is natural to examine whether this coupling is sensitive to charge doping.
Indeed, some compounds in this family, such as Rb$_{1-\delta}$V$_{2}$Te$_{2}$O \cite{ablimit_weak_2018,zhang_crystal-symmetry-paired_2025}, could exhibit intrinsic hole doping due to alkali metal deficiency.
% Systematic Trends with Alkali Metal Variation
\begin{figure}[h]
    \centerline{\includegraphics[scale=0.5]{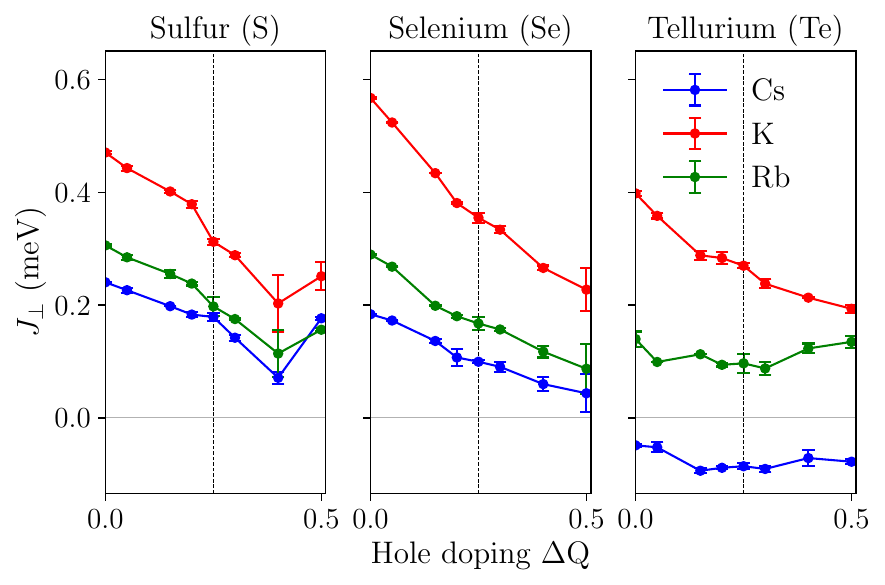}}
    \caption{\justifying Interlayer exchange coupling $J_{\perp}$ as a function of hole doping $\Delta Q$ (per f.u.). $\Delta Q=0$ corresponds to the stoichiometric composition. The error bars show the difference between the calculations performed with $26\times26\times7$ and $22\times22\times7$ $k$-point meshes.}.
    \label{fig:jdope}
\end{figure}

\begin{figure}[h]
    \centerline{\includegraphics[scale=0.5]{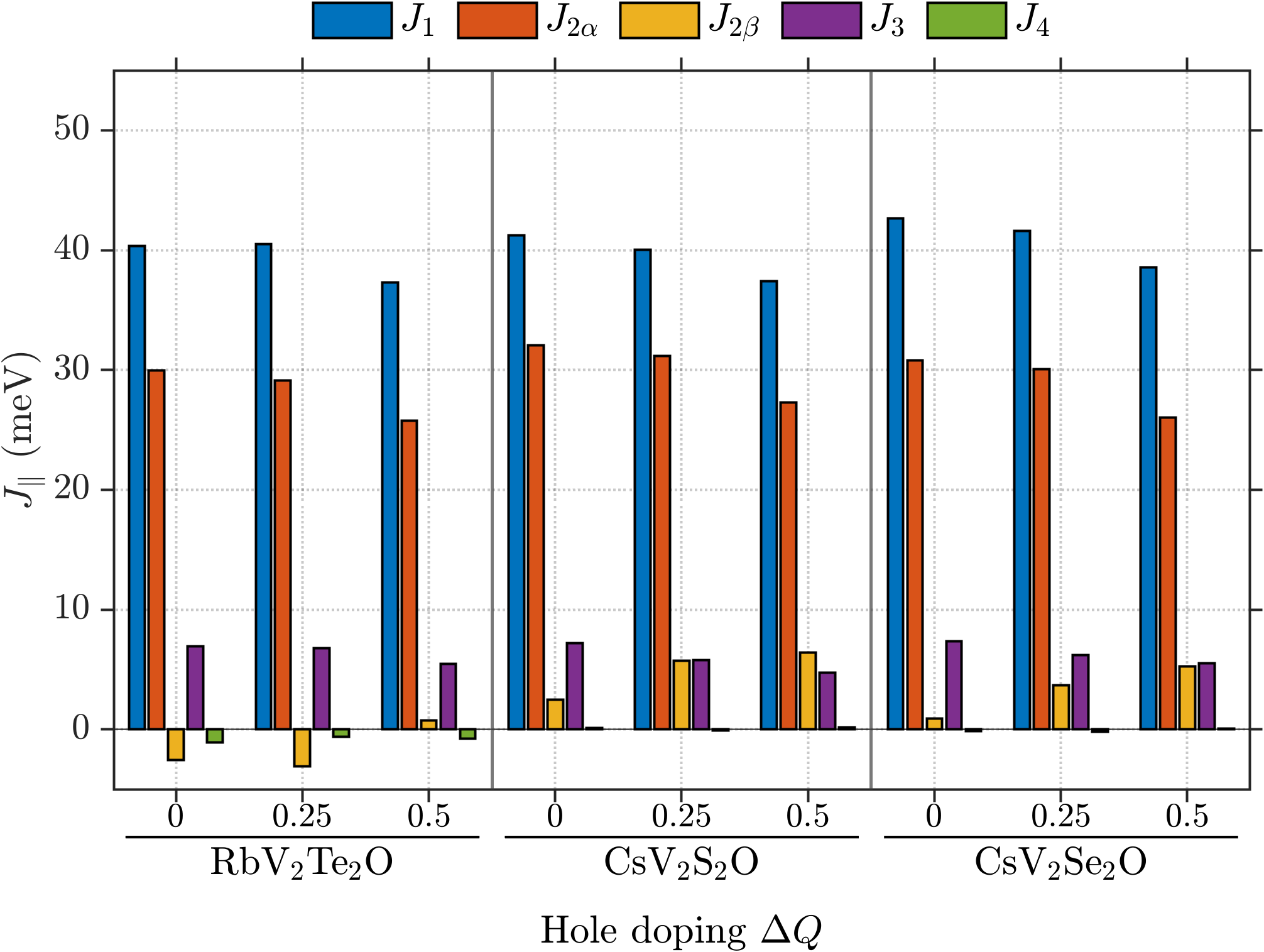}}
    \caption{\justifying 
    Dependence of the intralayer exchange parameters on hole doping $\Delta Q$ (holes per f.u.) for selected compounds.
}
    \label{fig:jparadope_omx}
\end{figure}

The hole-doping dependence of the interlayer exchange coupling $J_\perp$ 
is shown in Fig. \ref{fig:jdope}.

We observe monotonic decline in $J_{\perp}$  as the filler layer changes from K to Rb to Cs while the chalcogen element remains the same. 
This systematic trend is most likely due to the significant increase in the interlayer spacing as larger alkali metal cations
are incorporated into the structure. 

We also observe monotonic decline in $J_{\perp}$ under doping of up to 0.25 holes per formula unit.
K-based compounds (red lines in Fig. \ref{fig:jdope}) show the steepest decline but remain strongly positive. Rb-based compounds (green lines) also show weakening of $J_{\perp}$, but in tellurides this trend is much weaker. 
In Cs-based compounds with S and Se (blue lines), $J_\perp$ first decreases strongly and then tends to saturate, remaining positive up to 0.5 holes per f.u.
In \ch{CsV2Te2O}, which is the only case with $J_\perp<0$ at stoichiometry, hole doping has no significant effect on $J_\perp$. 
Thus, hole doping within a reasonable range does not change the sign of $J_\perp$ in any of the studied compounds.

Figure \ref{fig:jparadope_omx} shows that intralayer exchange parameters are rather insensitive to hole doping, implying that intralayer AM order is also stable under moderate deviations from stoichiometry.

\paragraph{Conclusions ---\label{sec:conclusion}}

Our systematic study of nine vanadium-based oxychalcogenide inverse Lieb-lattice compounds suggests that all except one of them have a conventional antiferromagnetic ground state with the doubled magnetic unit cell along the $c$ axis. The only exception is \ch{CsV2Te2O}, which is predicted to be AM in the bulk and appears to be the prime candidate for experimental verification. This conclusion is not affected by doping of up to 0.5 holes per formula unit.

\bibliographystyle{apsrev4-2}
\bibliography{main.bib} 
\end{document}